\documentclass[conference]{IEEEtran}
\IEEEoverridecommandlockouts

\usepackage{cite}
\usepackage{amsmath,amssymb,amsfonts}
\usepackage{algorithmic}
\usepackage{graphicx}
\usepackage{textcomp}
\usepackage{xcolor}

\usepackage{hyperref}
\usepackage{listings}
\usepackage{caption}
\usepackage{multirow}
\usepackage{float} 
\usepackage{rotating}
\definecolor{macrocolor}{RGB}{153, 0, 153}      
\definecolor{operatorcolor}{RGB}{0, 128, 128}   
\definecolor{keywordcolor}{RGB}{0, 0, 255}      
\definecolor{commentcolor}{RGB}{128, 128, 128}  
\definecolor{stringcolor}{RGB}{196, 26, 22}     

\definecolor{macrocolor}{RGB}{0, 0, 0}      
\definecolor{operatorcolor}{RGB}{0, 0, 0}   
\definecolor{keywordcolor}{RGB}{0, 0, 0}      
\definecolor{commentcolor}{RGB}{0, 0, 0}  
\definecolor{stringcolor}{RGB}{0, 0, 0}     

\lstdefinelanguage{Julia}{
    keywords={function, end, return, if, else, elseif, for, while, begin, let, do, try, catch, finally, break, continue, module, import, using, export, const, global, local, true, false, where},
    morekeywords=[2]{@inline, @kernel, @atomic, @cuda, @inbounds, @index},  
    sensitive=true,
    morecomment=[l]{\#},
    morecomment=[s]{\#=}{=\#},
    morestring=[b]",
    morestring=[b]',
}

\newcommand{\src}{\texttt{src}}
\newcommand{\dst}{\texttt{dst}}
\newcommand{\op}{\mathop{\mathrm{op}}}
\hypersetup{
    colorlinks=true,
    linkcolor=black,      
    citecolor=black,      
    urlcolor=blue,        
}
\def\BibTeX{{\rm B\kern-.05em{\sc i\kern-.025em b}\kern-.08em
    T\kern-.1667em\lower.7ex\hbox{E}\kern-.125emX}}
\begin{document}

\title{High-Performance Portable GPU Primitives for Arbitrary Types and Operators in Julia\\
}

\author{\IEEEauthorblockN{Emmanuel Pilliat}
\IEEEauthorblockA{\textit{Univ Rennes, Ensai, CNRS, CREST—UMR 9194, F-35000} \\
Rennes, France \\
emmanuel.pilliat@ensai.fr}
}

\maketitle

\begin{abstract}
Portable GPU frameworks such as Kokkos and RAJA reduce the burden of cross-architecture development but typically incur measurable overhead on fundamental parallel primitives relative to vendor-optimized libraries. We present \textbf{KernelForge.jl}, a Julia library that implements scan, mapreduce, and matrix--vector primitives through a two-layer portable architecture: \textbf{KernelIntrinsics.jl} provides backend-agnostic abstractions for warp-level shuffles, memory fences, and vectorized memory access, while KernelForge.jl builds high-performance algorithms exclusively on top of these interfaces. Evaluated on an NVIDIA A40 and an AMD MI300X, KernelForge.jl matches or exceeds CUB kernel execution time on scan and mapreduce on the A40, and matches cuBLAS throughput on matrix--vector operations across most tested configurations---demonstrating, as a proof of concept, that portable JIT-compiled abstractions can achieve vendor-level throughput without sacrificing generality.
\end{abstract}


\section{Introduction}
\label{sec:introduction}

Modern scientific computing, machine learning, and data analytics increasingly
rely on GPU acceleration to achieve necessary performance levels. However, the
GPU computing landscape has evolved from a single-vendor dominated ecosystem to
a heterogeneous environment with multiple competing architectures. NVIDIA's CUDA
platform, AMD's ROCm, Intel's oneAPI, and emerging architectures each offer
distinct hardware capabilities and programming models. This diversity creates a
fundamental tension: applications optimized for one architecture may fail to
compile or perform poorly on others.

The traditional approach of maintaining separate implementations for each
vendor's platform imposes substantial development and maintenance costs.
Research groups and software developers face a persistent trade-off: write
portable code that may underperform, or maintain multiple specialized
implementations that maximize performance but increase complexity. This tension
between performance and portability has been extensively documented in the HPC
community~\cite{sedova2018high, pennycook2019implications}.

Existing performance portability frameworks such as RAJA and Kokkos can incur
up to 100\% overhead compared to hand-optimized vendor
implementations~\cite{martineau2017assessing, artigues2020evaluation} due to
suboptimal code generation through abstraction layers. This performance gap is
particularly acute for fundamental parallel primitives---like scan, reduction, matrix-vector or vector matrix operations---that serve as building blocks for complex algorithms. These
primitives underpin a wide range of workloads, from sparse linear algebra
solvers to graph analytics and sorting routines; even modest overhead at this
level compounds through higher-level applications. A second limitation of
existing frameworks is their reliance on ahead-of-time compilation, which
complicates integration with dynamic, high-level languages such as Python and
Julia that increasingly dominate scientific computing workflows.

Within the Julia ecosystem specifically, GPU programming has matured rapidly
through packages like CUDA.jl~\cite{besard2019julia} and
KernelAbstractions.jl~\cite{Churavy_KernelAbstractions_jl}. Yet even Julia's
native GPU compiler introduces constant overhead versus vendor implementations
due to differences in low-level code generation~\cite{godoy2023evaluating}, and
portable GPU libraries in Julia have not closed this performance gap for core
primitives. Bridging this gap would benefit the growing Julia HPC and machine
learning community, which relies on these primitives as foundational
infrastructure.

We show that, contrary to common expectation, performance portability need not
require either performance compromise or mastery of complex low-level syntax.
Through careful design leveraging Julia's high-level abstractions,
metaprogramming capabilities, and just-in-time (JIT) compilation, it is
possible to achieve \emph{both} portability \emph{and} throughput competitive
with hand-optimized vendor libraries.

We present \textbf{KernelForge.jl}~\cite{kernelforge_not_anonimized}, a GPU library for Julia
that implements fundamental parallel primitives with throughput matching
vendor-optimized implementations. KernelForge.jl is architected around three
goals: \emph{performance}, matching vendor-optimized libraries on core
primitives; \emph{flexibility}, supporting arbitrary associative operators and
any Bitstype element---including for matrix--vector products, where vendor
libraries are restricted to standard numeric arithmetic; and
\emph{portability}, expressing all algorithms through backend-agnostic
abstractions with vendor-specific functionality isolated in a thin extension
layer. The current implementation targets NVIDIA GPUs via CUDA.jl and AMD GPUs
via AMDGPU.jl; extending to Intel oneAPI and other backends requires only
providing backend-specific implementations of the low-level intrinsics, with no
changes to the algorithmic layer. The principal contributions of this work are:

\begin{itemize}
    \item A two-layer architecture in which \textbf{KernelIntrinsics.jl}~\cite{kernelintrinsics_not_anonimized}
    provides low-level, portable intrinsics for vectorized memory access, configurable
    memory fences, and warp-level operations on arbitrary types, and
    \textbf{KernelForge.jl} implements high-performance parallel algorithms
    exclusively through KernelIntrinsics.jl and KernelAbstractions.jl~\cite{Churavy_KernelAbstractions_jl}.
    Backend-specific functionality is confined to KernelIntrinsics.jl extension
    modules via Julia's package extension mechanism and \texttt{@device\_override},
    ensuring that the algorithmic layer requires no modification when targeting
    new architectures.

    \item An empirical demonstration that JIT-compiled, high-level abstractions can match---and in some cases exceed---the throughput of hand-optimized vendor libraries on scan, reduce, and matrix--vector primitives, across problem sizes ranging from $10^6$ to $10^9$ elements, evaluated on an NVIDIA A40 and an AMD MI300X. On the A40, we report kernel-only execution time measured via CUDA Events alongside full pipeline time (memory allocation, host-side dispatch, and kernel execution); on the MI300X, we report full pipeline time only.

    \item A comparative evaluation against CUB, CUDA.jl, AcceleratedKernels.jl~\cite{nicusan2025acceleratedkernels},
and Kokkos (scan only), demonstrating that KernelForge.jl matches vendor-optimized
performance on scan, reduce, and matrix--vector primitives on the A40, and achieves
competitive throughput on the MI300X (vendor-optimized baselines for scan and mapreduce
are unavailable on that platform). KernelForge.jl
exposes a lightweight tuning mechanism based on Julia's native multiple dispatch,
allowing architecture-specific parameters to be selected at compile time with no
changes to the algorithmic layer.
\end{itemize}

We emphasize that KernelForge.jl is a proof of concept rather than a
production-ready library. The current implementation targets NVIDIA GPUs
fully; AMD support via AMDGPU.jl is functional but not yet accompanied by
vendor-optimized baselines for scan and mapreduce on the MI300X---rocPRIM
benchmarks are absent from this evaluation, so MI300X results for those
primitives demonstrate portability and correctness rather than substantiating
a no-portability-tax claim on that platform. Matrix--vector and vector--matrix
primitives are an exception: AMDGPU.jl dispatches \texttt{LinearAlgebra.mul!}
to \texttt{rocblas\_gemv} internally, providing an effective rocBLAS baseline
for those operations. Backends beyond CUDA and ROCm remain untested. The
primary contribution is architectural and empirical: demonstrating that the
portability--performance trade-off is not fundamental, and that a carefully
designed high-level abstraction layer can match vendor-optimized throughput
on core parallel primitives.

\section{Background}
\label{sec:background}

\subsection{GPU Architecture}
\label{sec:gpu_arch}
GPUs execute thousands of threads organized into \emph{warps}---the atomic
unit of execution---and \emph{blocks} of warps scheduled across Streaming
Multiprocessors (SMs)~\cite{nickolls2010gpu}. A memory hierarchy trades
capacity for speed: global memory is large but slow, shared memory is
block-scoped and fast, and registers are per-thread and fastest. Threads within
a warp can exchange data directly via shuffle instructions without touching
memory~\cite{de2019automatic}; warps within a block synchronize through shared
memory~\cite{hechtman2013exploring}; cross-block coordination requires global
memory fences or atomics~\cite{xiao2010inter,wang2019fast}.

Memory access performance depends critically on \emph{coalescing}---contiguous,
aligned warp accesses are serviced in a single transaction---and on issuing
wide loads (e.g., 128-bit \texttt{float4} in CUDA) to maximize bandwidth per
transaction~\cite{mei2016dissecting,rhu2013locality}.

These three mechanisms---warp-level primitives, memory fences, and vectorized
memory access---are the low-level capabilities required for vendor-competitive
parallel primitives, and form the focus of KernelIntrinsics.jl
(Section~\ref{sec:design_intrinsics}). Peak performance further requires tuning
block count and items-per-thread to hardware characteristics such as warp width
and L2 cache size that vary across architectures.

\subsection{Cross-Architecture GPU Programming in Julia}
\label{sec:julia_gpu}

Julia's GPU ecosystem is built on a layered stack. At the lowest level, vendor-specific packages---CUDA.jl for NVIDIA, AMDGPU.jl for AMD ROCm, oneAPI.jl for Intel, and Metal.jl for Apple---provide full access to each platform's programming model and vendor libraries. Above this, KernelAbstractions.jl~\cite{Churavy_KernelAbstractions_jl} provides a backend-agnostic kernel language: a single kernel definition can be compiled to any supported backend through Julia's multiple dispatch and JIT compilation.

KernelAbstractions.jl exposes a programming model in which the user defines kernels using the \texttt{@kernel} macro. Kernels are parameterized by a backend (e.g., \texttt{CUDABackend()}, \texttt{ROCBackend()}) and compiled at first invocation for the target device. The framework provides workgroup-level abstractions---including workgroup indices, local indices, and \texttt{@synchronize} barriers---that map to thread blocks, thread-local indices, and \texttt{\_\_syncthreads()} on NVIDIA, with analogous mappings on other backends. Shared memory is allocated via \texttt{@localmem}, which maps to \texttt{\_\_shared\_\_} memory on CUDA and local data share on AMD.

\begin{lstlisting}[caption={Copy kernel in KernelAbstractions.jl, from GPUArraysCore.jl. This kernel compiles to NVIDIA PTX, AMD GCN, Intel SPIR-V, or Apple AIR from a single source definition.}, label={lst:ka_copy}]
@kernel function copy_kernel!(dst, src)
    I = @index(Global)
    dst[I] = src[I]
end
\end{lstlisting}

Listing~\ref{lst:ka_copy} illustrates the simplicity of the programming model. This abstraction covers a substantial portion of the GPU programming surface, enabling portable kernels for many workloads. However, several capabilities essential for vendor-competitive performance on fundamental primitives are \emph{not} exposed:

\begin{itemize}
    \item \textbf{Warp-level shuffle operations.} There is no portable abstraction for \texttt{shfl\_sync}, \texttt{shfl\_down\_sync}, or their equivalents. Kernels requiring register-level data exchange within a warp---critical for fast reductions and scans---must fall back to vendor-specific intrinsics, breaking portability.

    \item \textbf{Ordered memory accesses.} Fine-grained memory fences (\texttt{threadfence()}, \texttt{threadfence\_block()}) that control the visibility ordering of memory operations across threads, blocks, or the device are not available. These are required for inter-block coordination protocols such as the decoupled look-back used in single-pass scan algorithms~\cite{merrill2016single}.

    \item \textbf{Vectorized loads and stores.} There is no mechanism to emit 64-bit or 128-bit load/store instructions. Achieving peak memory bandwidth in copy and bandwidth-bound kernels requires this capability.
\end{itemize}

These gaps motivate our \textbf{KernelIntrinsics.jl} layer (Section~\ref{sec:design_intrinsics}), which provides portable, backend-dispatched implementations of all three capabilities, enabling KernelForge.jl to build vendor-competitive algorithms without sacrificing portability.

\subsection{Parallel Primitives}
\label{sec:primitives}

Beyond the copy operation---which serves as the practical bandwidth ceiling against which other primitives are measured (cf.\ Figure~\ref{fig:vcopy})---several fundamental operations form the backbone of GPU computation, serving a role analogous to BLAS~\cite{blackford2002blas} in linear algebra. We focus on three: mapreduce, scan, and matrix--vector/vector--matrix products.

The \textbf{mapreduce} operation transforms each element via a mapping function $f: T \to S$, then combines all transformed values using a commutative and associative
operator $\op$ (commutativity is required here, in contrast to
\texttt{scan} which requires only associativity) to compute $\op(f(\src[1]), f(\src[2]), \ldots, f(\src[n]))$. Efficient GPU reduction proceeds hierarchically: threads first reduce within registers, then within warps using shuffle operations, then across warps within a block using shared memory, and finally across blocks. Implementations leveraging warp-level primitives achieve up to $7.8\times$ speedup over tree-based reductions through shared memory alone~\cite{harris2007scan, de2019automatic}.

The \textbf{scan} (prefix sum) operation relaxes the commutativity requirement, taking an associative operator $\op$ (not necessarily commutative) and computing a sequence of partial reductions: the output $\dst[i] = \op(\src[1], \src[2], \ldots, \src[i])$ contains the accumulated result up to position $i$ for all $1 \leq i \leq n$. Scan is among the most important parallel primitives: it enables stream compaction, radix sort, histogram computation, and sparse matrix operations~\cite{blelloch1990prefix, blelloch1989scans}. The state-of-the-art is the \emph{single-pass parallel prefix algorithm with decoupled look-back}~\cite{merrill2016single}, which achieves throughput approaching copy operations by requiring only approximately $2n$ data movement ($n$ inputs read, $n$ outputs written) and dissociating local computation from global prefix propagation latencies through strategic redundant work. This protocol requires inter-block communication through global memory with careful use of memory fences and status flags---precisely the capabilities absent from existing portable frameworks -- see Section~\ref{sec:julia_gpu}.

Finally, we treat \textbf{matrix--vector} and \textbf{vector--matrix} products as distinct primitives. Given an $n \times p$ matrix $A$, a vector $x \in T^n$, a mapping function $f: T \times T \to S$, and an associative reduction operator $\op$ over $S$, the matrix--vector product computes a vector $y \in S^p$ where $y[j] = \op_{i=1}^{n} f(x[i],\, A[i,j])$, reducing over the rows of $A$. The vector--matrix product symmetrically computes $z \in S^n$ where $z[i] = \op_{j=1}^{p} f(A[i,j],\, x[j])$, reducing over the columns. Setting $f = \times$ and $\op = +$ recovers the standard BLAS GEMV operation, but the generalized formulation supports arbitrary algebraic structures---for instance, tropical semirings (where $f = +$ and $\op = \min$) used in shortest-path algorithms, or log-space operations for numerical stability. These two operations are \emph{not} symmetric on GPUs: because arrays are stored in column-major order, matrix--vector products access $A$ with stride-one (coalesced) reads along columns, while vector--matrix products must read along rows, resulting in strided access patterns. Achieving high throughput for both orientations requires distinct kernel strategies, as we describe in Section~\ref{subsec:matvec}.

\section{Related Work}
\label{sec:related}

\subsection{C++ Performance Portability Frameworks}

Performance portability in GPU computing has been addressed through various
C++ frameworks, each embodying different design philosophies. RAJA~\cite{hornung2014raja}
achieves portability by decoupling algorithm implementation from execution
strategy through execution policies. Kokkos~\cite{edwards2014kokkos} offers a
more encompassing framework that abstracts both execution and memory spaces.
Additional frameworks pursuing similar goals include OpenACC, OpenMP, and SYCL.
Despite their differences, these frameworks share a fundamental constraint:
their dependence on ahead-of-time compilation hinders integration with dynamic
languages such as Python and Julia.

Performance studies reveal a consistent pattern: portable frameworks typically
incur overhead compared to hand-optimized vendor implementations. Martineau
et al.~\cite{martineau2017assessing} found 5--30\% penalties versus
architecture-specific code; Artigues et al.~\cite{artigues2020evaluation}
reported 2--3$\times$ slowdowns on NVIDIA V100s compared to native CUDA.
Davis et al.~\cite{davis2025taking} confirmed that Kokkos performs moderately
worse than CUDA across five proxy applications on V100, A100, and H100 GPUs.
Sedova et al.~\cite{sedova2018high} concluded that non-portable optimizations
are essential to production molecular dynamics codes. These studies establish
the conventional wisdom that a portability tax is unavoidable when using
abstraction layers.

\subsection{Vendor-Optimized Libraries}

For 1-D parallel primitives (copy, reduce, scan), NVIDIA's CUB~\cite{merrill2015cub} and Thrust libraries provide highly optimized CUDA implementations, achieving near-peak performance through careful exploitation of warp-level operations, shared memory banking patterns, and instruction-level parallelism. AMD's rocPRIM and Intel's oneDPL offer similar functionality for their respective platforms. The existence of multiple vendor-specific libraries with incompatible APIs illustrates the fundamental challenge: achieving peak performance currently requires separate implementations per vendor.

For matrix--vector operations, vendor libraries such as cuBLAS, rocBLAS, and oneMKL provide optimized GEMV routines, but only for standard numeric types with fixed arithmetic ($f = \times$, $\oplus = +$). Applications requiring non-standard algebraic structures---such as tropical semirings for shortest-path problems or log-space operations for numerical stability---cannot use these libraries and must resort to custom kernels, forfeiting vendor-level optimization. KernelForge.jl's generalized formulation subsumes standard GEMV while extending it to arbitrary types and operators, without sacrificing performance on the standard numerical case.

\subsection{Julia GPU Ecosystem}

The Julia GPU ecosystem provides comprehensive GPU support through several packages. CUDA.jl offers both high-level array abstractions and low-level kernel programming for NVIDIA GPUs. KernelAbstractions.jl~\cite{Churavy_KernelAbstractions_jl} extends portability to multiple backends (CUDA, ROCm, oneAPI, Metal) through unified kernel syntax and backend dispatch. AcceleratedKernels.jl~\cite{nicusan2025acceleratedkernels} builds atop KernelAbstractions.jl to provide portable implementations of common operations. However, existing Julia GPU libraries prioritize generality over optimization: CUDA.jl's implementations favor simplicity, while AcceleratedKernels.jl focuses on broad portability rather than matching vendor library performance.

\subsection{Our Contribution}

KernelForge.jl differs from the frameworks above in two respects. First,
unlike C++ portability layers that compile ahead of time, it exploits
Julia's JIT compilation to generate code specialized to the concrete data
type, operator, and problem size at the call site---eliminating the
abstraction overhead that accounts for the 5--30\% portability tax
reported in existing studies. Second, unlike Julia GPU libraries that
prioritize API breadth, KernelForge.jl targets a narrow set of
primitives---scan, mapreduce, and matrix--vector products---and optimizes
them through a two-layer architecture: KernelIntrinsics.jl confines all
vendor-specific functionality (warp shuffles, memory fences, vectorized
loads) to a thin extension layer, while the algorithmic layer builds
exclusively on these portable abstractions. On the A40, this design
matches CUB and cuBLAS throughput on all three primitives; on the MI300X,
it delivers competitive performance on mapreduce and scan relative to the
available Julia baselines, and matches rocBLAS on matrix--vector
operations for wide-matrix configurations.

\section{Design of Cross Architecture Kernel Intrinsics}\label{sec:design_intrinsics}

\subsection{Shuffle Operations}\label{sec:shuffle}

A shuffle operation within a warp enables a given lane to read a value directly from the register of another lane, as determined by a source index or mask. Vendor implementations, however, define shuffle intrinsics only for a narrow set of types---typically 32-bit primitives such as \texttt{Float32} and \texttt{Int32}.

CUDA.jl partially addresses this limitation through a recursion mechanism that extends shuffling to types like \texttt{Int64} or complex numbers, but support for more composite types such as tuples or quaternions remains unavailable.

KernelIntrinsics.jl generalizes this approach by leveraging Julia's \texttt{@generated} functions to recursively decompose any composite type at compile time. The key observation is that once shuffle intrinsics are defined for a single 32-bit primitive (e.g., \texttt{UInt32}), any composite type can be supported by decomposing it into its constituent primitive fields and shuffling each independently. We formalize this through a recursive definition: a \emph{Bitstype} is either a concrete primitive of at most 64 bits, a tuple of Bitstypes, or a struct whose fields are all Bitstypes. Using \texttt{@generated} functions, the compiler statically unrolls the recursion over struct fields and tuple elements, producing specialized, zero-overhead code for each concrete type. KernelIntrinsics.jl implements shuffle operations over arbitrary Bitstypes under this definition, fully abstracting away vendor-specific type restrictions with no runtime cost.

This generalized shuffle mechanism enables fast warp-level communication for arbitrarily complex data types entirely through registers, without resorting to shared memory. KernelForge.jl builds on this capability to expose a high-level interface: the user can simply write \texttt{mapreduce(f, op, src)} or \texttt{scan(f, op, src)} and obtain correct, efficient results even when the elements of \texttt{src} are complex structures, tuples, or nested compositions thereof.

\subsection{Ordered Memory Access}
\label{subsec:ordered-memory-access}

When a thread stores an element, the write is not instantly visible to all
other threads. The value is first written to the L1 cache, where it is
visible only to threads within the same workgroup. It is then propagated
to the L2 cache, which is shared across workgroups. This creates a
potential communication hazard: a thread in a different workgroup may load
a stale value from its own L1 cache before the updated value has arrived
in the L2 cache.

Consider a producer-consumer pattern where one workgroup writes data to
global memory and then sets a flag to signal completion. A release semantic
on the flag store guarantees that all prior writes are visible to any thread
that subsequently observes the flag. A matching acquire semantic on the
corresponding load guarantees that all subsequent reads observe
up-to-date values. Together, release--acquire pairs establish a
happens-before relationship between workgroups without requiring a full
system-wide fence.

KernelIntrinsics.jl exposes this pattern through the \texttt{@access}
macro: \texttt{@access flag[i] = 0x01} emits a release store, while
\texttt{x = @access flag[i]} emits an acquire load. On NVIDIA GPUs, these
lower directly to PTX memory ordering annotations --- for instance,
\texttt{st.release.gpu.global} for the store and \texttt{fence.acq\_rel.gpu}
for a GPU-scoped fence. On AMD GPUs, the equivalent semantics are expressed
through GCN scope bits (\texttt{sc1}/\texttt{sc0}) combined with explicit
cache-flush and invalidation instructions (\texttt{buffer\_wbl2},
\texttt{buffer\_inv}). Both lowerings are verified at the assembly level in
KernelIntrinsics.jl's test suite, confirming that Julia's compilation
pipeline introduces no unintended fences or reorderings. For backends where
fine-grained ordering annotations are unavailable, \texttt{@access} falls
back to a full memory fence, which is less precise but preserves the same
correctness guarantees with no changes to the algorithmic layer.

\subsection{Vectorized Memory Access}\label{subsec:vectorized}

\begin{figure}[htbp]
\centerline{\includegraphics[scale=0.34]{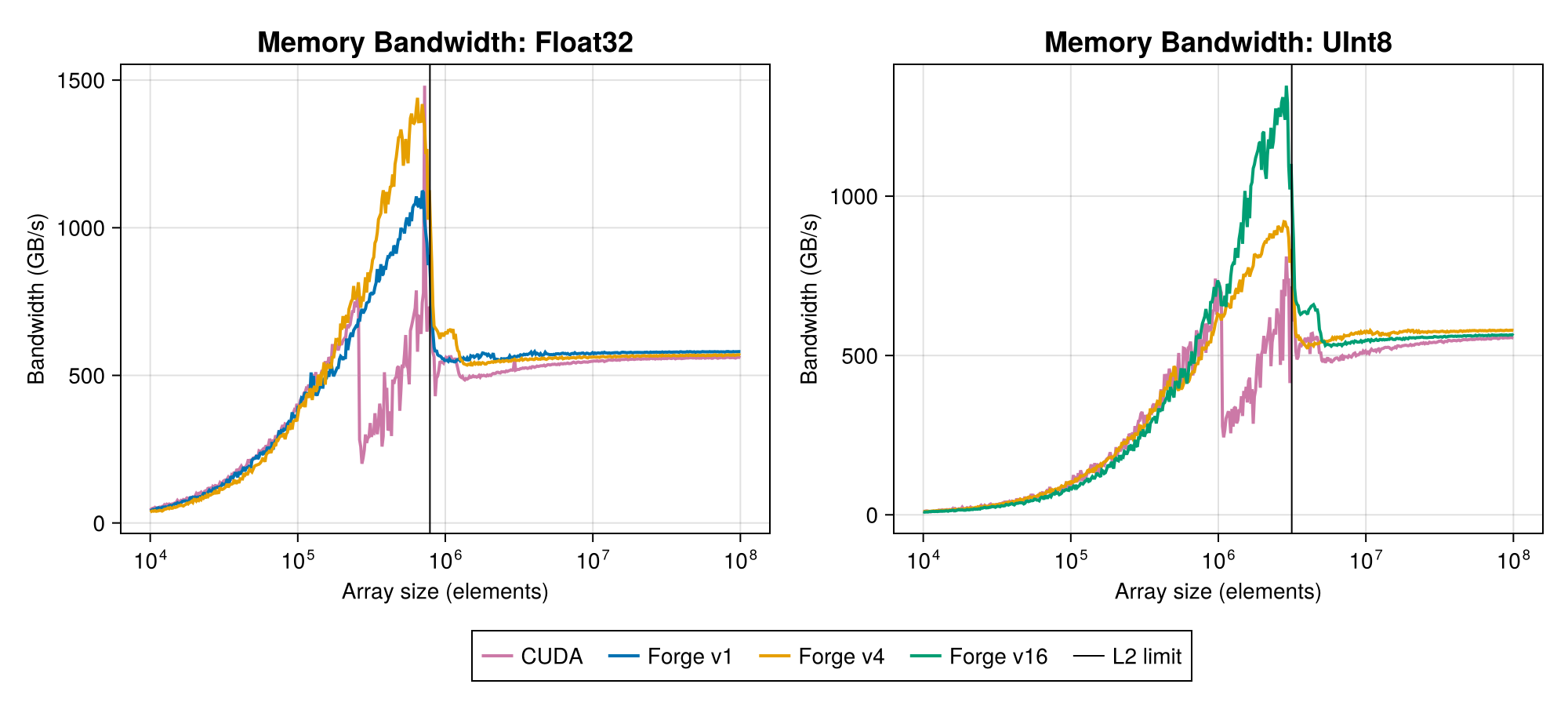}}
\caption{\textbf{Bandwidth (GB/s) for vectorized copy as a function of problem size.} Empirical bandwidth measured via kernel timing with \texttt{CUDA.@profile}, shown for CUDA.jl (which internally calls \texttt{libcuda}) and for KernelForge.jl with 1, 4, and 8 items per thread. The dashed vertical line indicates the L2 cache size divided by $2 \times \texttt{sizeof(element)}$. Peak bandwidth is achieved with 128-bit loads, where KernelForge.jl outperforms CUDA.jl.}
\label{fig:vcopy}
\end{figure}

GPU kernel performance is limited by either compute throughput or memory bandwidth~\cite{williams2009roofline, yang2020hierarchical}. For large problems that saturate memory bandwidth, vectorized loads---where each thread issues wide transactions (e.g., 128-bit loads of four \texttt{Float32} values) rather than scalar ones---are essential to maximize bandwidth utilization, particularly for problems that fit within the L2 cache.
KernelForge.jl defines its kernels with a static parameter \texttt{Nitem} that controls how many elements each thread processes, specialized at compile time via Julia's \texttt{Val{}} type with zero overhead. The optimal \texttt{Nitem} is not necessarily dictated by the vector load width: the \texttt{scan} kernel uses 16 \texttt{Float32} values per thread, since processing more elements sequentially amortizes synchronization cost across lanes and warps. Figure~\ref{fig:vcopy} illustrates this on a copy kernel, where 128-bit loads maximize bandwidth and KernelForge.jl matches or exceeds CUDA.jl's internal \texttt{libcuda} implementation.

Vectorized loads require contiguous memory access. For strided subarrays, KernelForge.jl falls back to loading individual elements into a tuple, forgoing vectorized loads but preserving the multi-item-per-thread structure and its synchronization benefits.

\subsection{Alignment Constraints}

Vectorized loads and stores provide substantial performance gains for problems that fit within the L2 cache, but they impose strict alignment requirements. The PTX ISA mandates that the address of any memory access be aligned to a multiple of the access size~\cite{ptx_isa}. For instance, a \texttt{ld.v4.f32} instruction loads 16 bytes and therefore requires the starting address to be 16-byte aligned. In practice, this means that a thread issuing a 128-bit load of \texttt{Float32} elements can only begin reading at element indices that are multiples of 4 (using 0-based indexing). A load starting at index 2 or 3 would violate this constraint and result in undefined behavior or a hardware fault.

More generally, for a vectorized load of \texttt{Nitem} elements of size $s$ bytes, the byte address must be a multiple of $\texttt{Nitem} \times s$. This creates a critical issue for matrix operations: since a matrix may have a number of rows $n$ that is not a multiple of \texttt{Nitem}, the starting address of the second column (and subsequent columns) is not guaranteed to satisfy the alignment constraint, even if the first column is properly aligned.

A natural solution is to decompose each misaligned load at runtime into a sequence of smaller, properly aligned loads. For example, if \texttt{Nitem}~$= 4$ and the starting index $i$ has a misalignment of 2 (i.e., $(i-1) \bmod \texttt{Nitem} = 2$), the load can be split into a scalar load at $i$, a vectorized 2-element load at $i+1$, and a scalar load at $i+3$. KernelIntrinsics.jl formalizes this through the function \texttt{vload\_pattern}, which is a \texttt{@generated} function that emits an optimal load sequence for a statically known alignment pattern, expressed as a tuple of integers summing to \texttt{Nitem} (e.g., $(1, 2, 1)$ in the example above). The outer \texttt{vload} function then dispatches at runtime to the appropriate \texttt{vload\_pattern} specialization via a compile-time-generated switch table, so that only the branch selection occurs at runtime while each load sequence itself is fully specialized.

Crucially, this mechanism is entirely hidden from the user. As shown in Listing~\ref{lst:vcopy}, a vectorized copy kernel in KernelForge.jl requires only a call to \texttt{vload} and \texttt{vstore!}---alignment handling, pattern decomposition, and specialization are managed transparently by the library and at compile time using multiple dispatch. 

\begin{figure}[t]
\begin{lstlisting}[language=Julia, caption={Vectorized copy kernel with $4$ elements per thread in KernelForge.jl. Alignment handling is managed transparently by \texttt{vload} and \texttt{vstore!}.}, label={lst:vcopy}]
@kernel function vcopy!(dst, src)
    I = @index(Global)
    vals = vload(src, I, 4)
    vstore!(dst, I, vals)
    # copy remaining elts if I == ndrange
end
\end{lstlisting}
\end{figure}

\section{Design of Fast Parallel Primitives}
\label{sec:implementation}

\subsection{Mapreduce}
\label{subsec:mapreduce}

The \texttt{mapreduce} primitive (Section~\ref{sec:primitives}) is implemented
with a fixed grid of blocks (e.g., 100 blocks of 256 threads, tuned per
architecture). Each thread strides across the input array with stride equal to
the total thread count, accumulating a partial result in registers. This
fixed-grid strategy processes arbitrarily large arrays without growing the
launch configuration.

Within each block, partial results are reduced hierarchically: first across
threads within each warp using shuffle operations, then across warps via shared
memory, yielding one partial value per block. Warp-level shuffles are provided
by KernelIntrinsics.jl (Section~\ref{sec:shuffle}) and support arbitrary
composite types at no runtime cost.

Inter-block aggregation is handled without a second kernel launch. Each block
writes its partial result to global memory and raises a \texttt{UInt8}
completion flag, initialized to zero before launch. A release-semantic store,
issued via the \texttt{@access} macro, guarantees that the partial result is
visible to all other blocks before the flag is observed. A designated block
spins on all flags using acquire-semantic loads---also via \texttt{@access}---
and performs the final reduction once all partial results are confirmed visible.
This single-launch design avoids the two-kernel sequence used by CUDA.jl.

The implementation selects among three paths based on input dimensionality.
One-dimensional arrays use the strategy above. Two-dimensional arrays are
dispatched to the matrix-vector and vector-matrix kernels
(Section~\ref{subsec:matvec}), which are already optimized for coalesced access
and warp-level parallelism along both axes. Higher-dimensional arrays and
arbitrary iterators fall through to a general reduction kernel, at a modest
performance cost relative to the specialized paths.

\subsection{Scan}
\label{subsec:scan}

The \texttt{scan} primitive computes an inclusive or exclusive prefix
reduction over an input array using any associative operator. It shares
the same building blocks as \texttt{mapreduce}---vectorized loads,
warp-level shuffles, and release/acquire flag synchronization via
\texttt{@access}---and achieves optimal single-pass throughput via the
decoupled lookback algorithm of Merrill and
Garland~\cite{merrill2016single}.

The input is partitioned into tiles of $256 \times \texttt{Nitem}$
elements, one tile per thread block. Each thread loads \texttt{Nitem}
elements using \texttt{vload} (Section~\ref{subsec:vectorized}),
accumulates a local prefix entirely in registers, then participates in a
warp-shuffle reduction and a shared-memory exchange to compute the tile's
aggregate. Global memory is therefore read exactly once per element. Upon
completion, a designated thread stores the tile aggregate and raises a
\emph{partial} status flag via a release store (\texttt{@access flag[i] = PARTIAL}).

Global prefix propagation proceeds via the lookback phase. A single warp
scans backwards over the preceding 32 tiles (64 on AMD wavefront-64
hardware), spinning on each status flag via acquire loads
(\texttt{@access flag[j]}) until at least a partial flag is visible.
If all 32 (or 64) preceding tiles have raised only partial flags, the
warp reduces their aggregates via shuffle-based reduction and continues
looking back at the next group of tiles. As soon as the warp encounters
a tile with a \emph{prefix} flag---indicating that the complete inclusive
prefix through that tile is already available---it can immediately
finalize the current tile's inclusive prefix without inspecting any
earlier tiles. The finalized prefix is written to global memory and the
tile raises its own prefix flag via a release store, unblocking all
subsequent tiles. The correctness of this scheme relies on the
release--acquire ordering established by \texttt{@access}
(Section~\ref{subsec:ordered-memory-access}): without it, a tile could
observe a raised flag while the associated aggregate value remains stale
in a remote L1 cache.

Once the inclusive prefix is known, each thread computes its final output
values entirely in registers and writes them back to the destination array
using vectorized stores, so global memory is written exactly once per
element.

The resulting implementation is correct for any associative operator,
including noncommutative cases such as quaternion multiplication, and
achieves throughput competitive with CUB on standard types
(Section~\ref{sec:performance}).
\subsection{Matrix-Vector and Vector-Matrix Products}
\label{subsec:matvec}

Matrix-vector and vector-matrix products are central to neural network
inference and training, and are typically served by highly optimized proprietary
libraries such as cuBLAS. These libraries, however, are vendor-locked and
restricted to standard numeric arithmetic. KernelForge.jl provides
open-source implementations that match cuBLAS throughput on CUDA for standard
types, while supporting arbitrary element types and associative operator pairs.
This generality enables use cases beyond the standard $(\times, +)$ semiring---
such as tropical semiring operations for shortest-path computations or
log-space accumulation for numerical stability---without sacrificing performance
on the standard numerical case.

\begin{figure}[htbp]
\centerline{\includegraphics[scale=0.2]{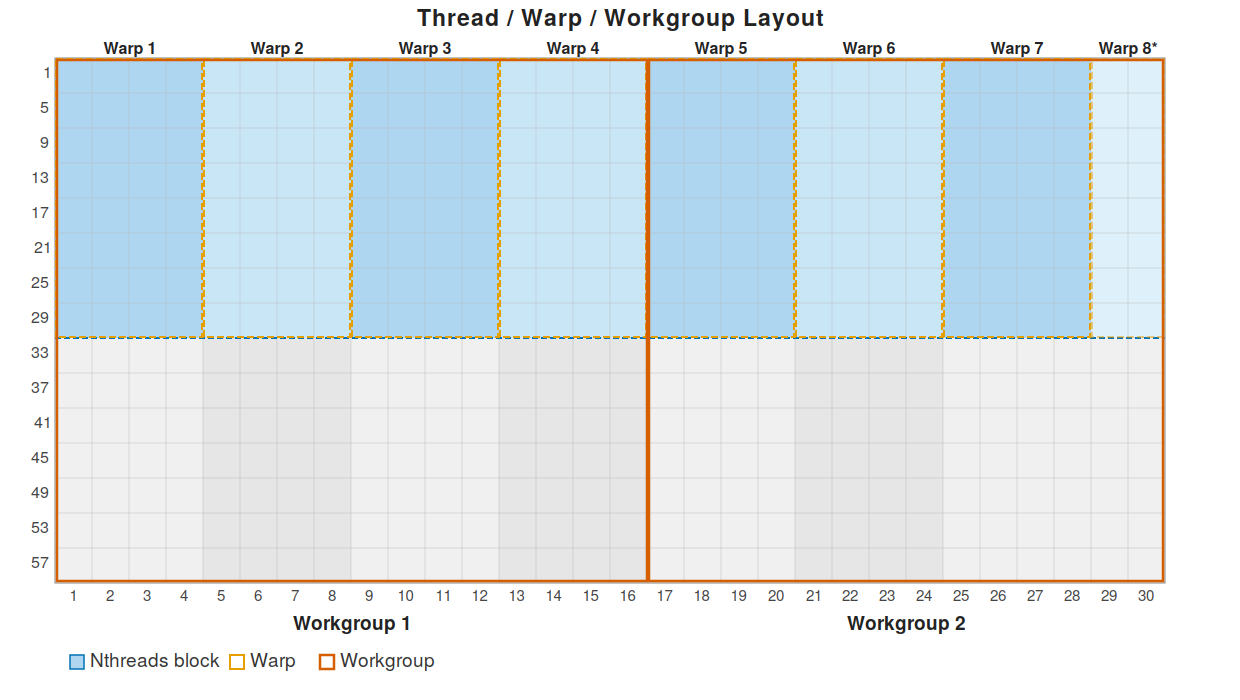}}
\caption{\textbf{Vector-matrix thread organization for a wide, short matrix.}
The $x$-axis represents columns and the $y$-axis represents rows. Each warp
(dashed yellow boundary) is assigned 4 consecutive columns; threads stride
vertically across rows, with the blue and grey regions corresponding to the
first and second row strides, respectively, each thread loading 4 elements per
stride. Each workgroup of 128 threads (solid orange boundary) thus covers 16
columns. This layout maintains coalesced memory access while keeping all
threads occupied across multiple strides.}
\label{fig:vecmat_layout}
\end{figure}

Because matrices are stored in column-major order, horizontal reductions
(matrix-vector products) and vertical reductions (vector-matrix products) access
memory along different axes and require distinct kernels with different
coalescing strategies (Section~\ref{sec:primitives}). The optimal thread
organization also depends on matrix shape. For tall, narrow matrices, each
column resembles an independent 1-D reduction: the same fixed-grid block
striding used in \texttt{mapreduce} applies directly, with a small number of
blocks (default: 100, tunable per architecture) assigned per column. For wide,
short matrices, work must instead be distributed across both dimensions to keep
all threads occupied; Figure~\ref{fig:vecmat_layout} illustrates this layout
for the vector-matrix case. KernelForge.jl selects the appropriate strategy at
kernel launch based on matrix shape, with dispatch resolved statically via
Julia's \texttt{Val} type so that each path is independently specialized by the
compiler with no runtime overhead.

\section{Correctness Validation}

Implementing fast parallel primitives with flexible operators introduces
significant testing complexity, as two distinct categories of errors arise
naturally during development.

The first category concerns algorithmic correctness: verifying results on large
arrays, handling edge cases such as sizes of 31 or 33 elements that straddle
warp boundaries, and ensuring the absence of race conditions. Rather than
formal proofs, which are beyond the scope of this work, we adopt a
comprehensive empirical approach, testing across a wide range of array sizes
and scalar types.

The second category concerns compilation validity. Julia compiles through LLVM
before generating PTX or GCN code, and any instruction that LLVM cannot
correctly lower produces errors that are often difficult to diagnose---ranging
from explicit IR failures to silent device-level faults that manifest as
spurious out-of-bounds accesses or misaligned memory operations, either of
which can crash the Julia session without a clear error message. Such failures
can be triggered by user-provided operators, custom structs with non-standard
alignment, or array views with non-trivial memory layouts. Our test
suite~\cite{kernelforge_not_anonimized} exercises all three sources: custom operators,
deliberately misaligned structures, and both contiguous and non-contiguous
array views.

Critically, the \emph{same} test suite runs unmodified on NVIDIA hardware, with competitive results on AMD, validating that KernelForge.jl produces correct results on the A40
(CUDA) and the MI300X (ROCm) without any backend-specific test paths. This
cross-architecture validation confirms that the portability guarantees of the
KernelIntrinsics.jl abstraction layer hold in practice: correctness is not
incidental to a single backend but is a property of the shared algorithmic layer.

\section{Performance Evaluation}\label{sec:performance}
We evaluate KernelForge.jl against several reference implementations:
CUDA.jl, which provides open-source implementations of \texttt{mapreduce}
and \texttt{accumulate}; AcceleratedKernels.jl, a cross-architecture Julia
library; and cuBLAS, the reference for matrix-vector and vector-matrix
operations (called internally by CUDA.jl for these operations on concrete
types). We additionally benchmark CUB directly using an \texttt{nvcc}
benchmark available in the \texttt{perf} folder of~\cite{kernelforge_not_anonimized}.
For the scan primitive, we further include
Kokkos~\cite{edwards2014kokkos,trott2021kokkos}, a widely used C++
performance portability framework, as an additional point of comparison.

\subsection{Experimental Setup}

\paragraph{Hardware Platforms}
Experiments are conducted on two GPU platforms. The \textbf{NVIDIA A40} is
an Ampere-generation GPU with 48\,GB of GDDR6 memory and 696\,GB/s peak
memory bandwidth, running CUDA 12.8 on Ubuntu 22.04. The \textbf{AMD
MI300X} is a CDNA3-generation GPU with 192\,GB of HBM3 memory and
5.3\,TB/s peak memory bandwidth, running ROCm on Ubuntu 24.04. These two
platforms represent distinct points in the GPU landscape: a mainstream
data-center GPU with mature CUDA tooling, and a high-memory-bandwidth
accelerator with a less mature software ecosystem.

\paragraph{Measurement Methodology}
Performance is evaluated using two complementary metrics.

\textit{GPU execution time} measures on-device kernel execution excluding
launch overhead, reflecting the quality of generated code and
representative of production scenarios where kernels are invoked
repeatedly. For Julia libraries on CUDA this is obtained via the
\texttt{CUDA.@profile} macro, which relies on CUDA Events internally; for
CUB we use CUDA Events directly. On the AMD backend, timing uses HIP events
(\texttt{AMDGPU.HIP.HIPEvent}), which measure device-side elapsed time
including kernel launch overhead. The \texttt{CUDA.@profile} macro isolates
individual kernel segments with higher precision than raw HIP events;
however, equivalent profiling infrastructure is not yet mature in
AMDGPU.jl. Allocation is excluded for all libraries that expose a
pre-allocation interface; on the AMD backend this is particularly
consequential, as \texttt{AMDGPU.jl} device memory allocation incurs
substantially higher overhead than its CUDA counterpart---benchmarks for
libraries without a pre-allocation interface should therefore be interpreted
as upper bounds on achievable kernel time rather than direct comparisons.
This metric is reported on both platforms.

\textit{End-to-end pipeline time} measures total wall-clock time including
CPU-side overhead: temporary memory allocation, kernel launch, result
transfer from device to host, and CUDA.jl API calls. First-compilation
costs due to Julia's JIT model~\cite{besard2018effective} are excluded, as
they amortize across repeated invocations. This metric is reported on the
A40 only; on ROCm, CPU-side timing exhibits substantially higher variance
and is not reported. Note that end-to-end pipeline is not directly comparable to CUB
timings, which measure only on-device execution.

\paragraph{Tuning Parameters.}
Achieving performance competitive with vendor-optimized libraries requires
careful tuning of parameters such as the number of items processed per
thread, the number of threads per block, and the number of blocks per
streaming multiprocessor. These parameters depend on hardware-specific
characteristics such as L2 cache size, register file capacity, and memory
bandwidth (see Figure~\ref{fig:vcopy}), and each must be chosen from a
discrete set of powers of two, resulting in a combinatorial tuning space.
This is particularly pronounced for the matrix-vector and vector-matrix
kernels, which expose several interdependent parameters controlling the
partitioning of work across threads, warps, and blocks.

KernelForge.jl addresses this through an architecture dispatch hierarchy
(\texttt{A40 <: Ampere <: AbstractArch}) that selects default parameters
at compile time via Julia's multiple dispatch, analogously to CUB's
per-PTX-version tuning policies~\cite{cub_tuning}. The parameters used
throughout Section~\ref{sec:results_a40} were manually tuned to match
vendor performance on the A40; matching performance on a new architecture
would require a comparable tuning effort for that target. Automated
tuning would reduce this burden and is a natural
direction for future work, but is outside the scope of this paper.

\subsection{Results on A40}\label{sec:results_a40}

\begin{figure}[htbp]
\centerline{\includegraphics[scale=0.35]{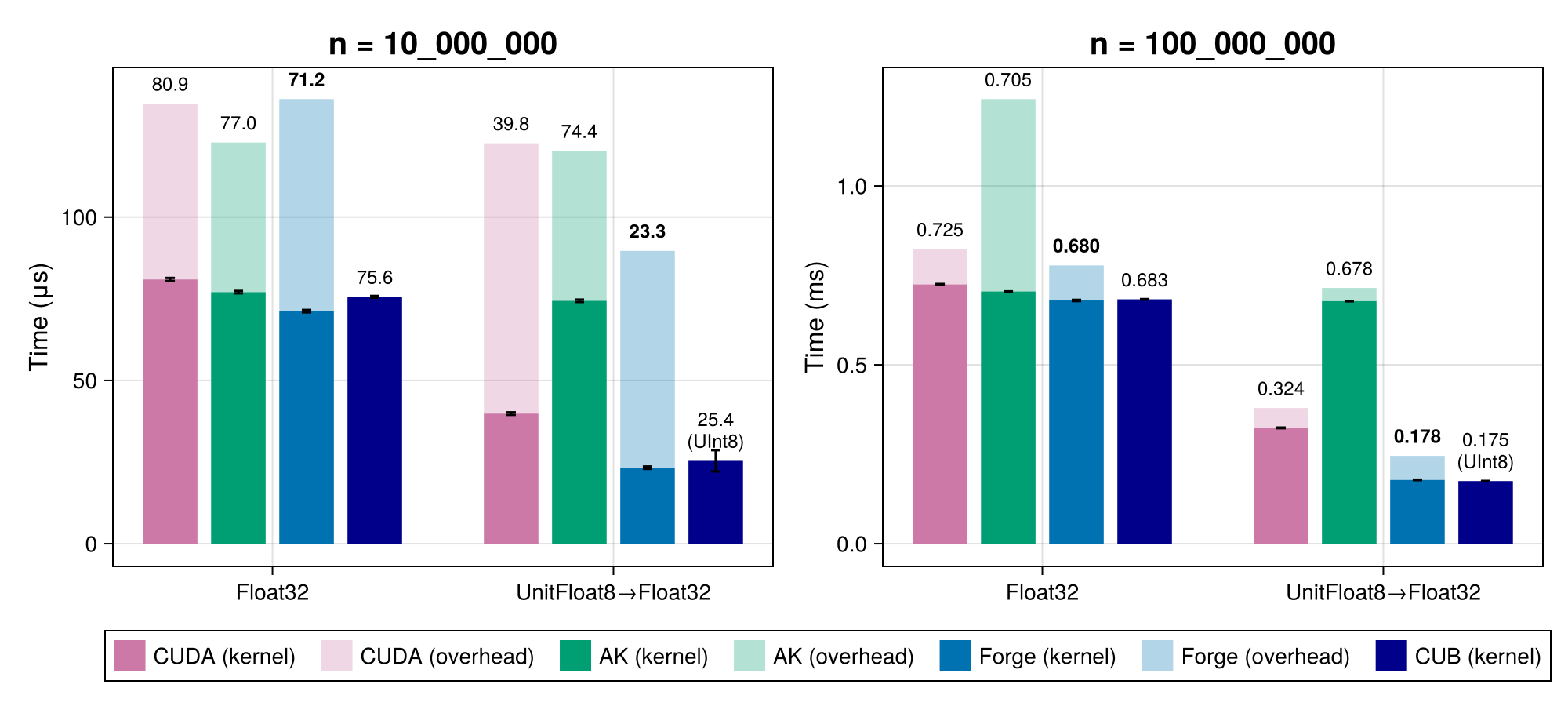}}

\caption{\textbf{Reduction Benchmark Across Implementations (A40).} Comparison of four implementations for the parallel sum operation: CUDA.jl, AcceleratedKernels.jl, KernelForge.jl, and CUB (compiled with \texttt{nvcc}). Results are shown for two input sizes: $n = 10^7$ (left) and $n = 10^8$ (right), and two data types: Float32 and UInt8. Dark bars show kernel execution time; light bars include launch overhead. Error bars indicate variability across runs. Since UnitFloat8 is a Julia-specific type, the CUB benchmark uses a dummy UInt8 summation for reference.}
\label{fig:mapreduce_perfs_A40}
\end{figure}

\paragraph{Mapreduce}
Figure~\ref{fig:mapreduce_perfs_A40} presents benchmark results for the
\texttt{mapreduce} primitive, comparing KernelForge.jl against CUDA.jl,
AcceleratedKernels.jl, and CUB for input sizes $n \in \{10^7, 10^8\}$
and two data types: \texttt{Float32} and a custom \texttt{UnitFloat8} Julia type
(\texttt{UInt8} for CUB).

On \texttt{Float32}, all implementations achieve comparable throughput at
these sizes. KernelForge.jl matches CUB while slightly outperforming
CUDA.jl and AcceleratedKernels.jl. The single-launch flag-based design
(Section~\ref{subsec:mapreduce}) yields a more pronounced advantage at
smaller sizes where inter-block synchronization overhead dominates, as
shown in Table~\ref{tab:mapreduce_benchmark}.

We also evaluate performance on a custom 8-bit type. \texttt{UnitFloat8}
encodes values in $[-1, 1]$ using 256 evenly spaced levels; elements are
promoted to \texttt{Float32} before summation to avoid overflow. Since
\texttt{UnitFloat8} is implemented in KernelForge.jl and has no CUB
equivalent, the CUB benchmark uses raw \texttt{UInt8} summation as a
reference lower bound. Correctness is validated by checking that the sign
of the result matches a reference \texttt{Float64} computation on CPU.

KernelForge.jl matches CUB on raw \texttt{UInt8}, achieving a $1.8\times$
speedup over CUDA.jl and a $3.8\times$ speedup over AcceleratedKernels.jl
at $n = 10^8$. Notably, this promotion incurs no measurable overhead: the
kernel is memory-bound at this scale, so the additional arithmetic is fully
hidden behind memory latency. This is enabled by vectorized memory access:
loading multiple 8-bit elements per transaction brings effective bandwidth
in line with 32-bit operations.

\begin{figure}[htbp]
\centerline{\includegraphics[scale=0.35]{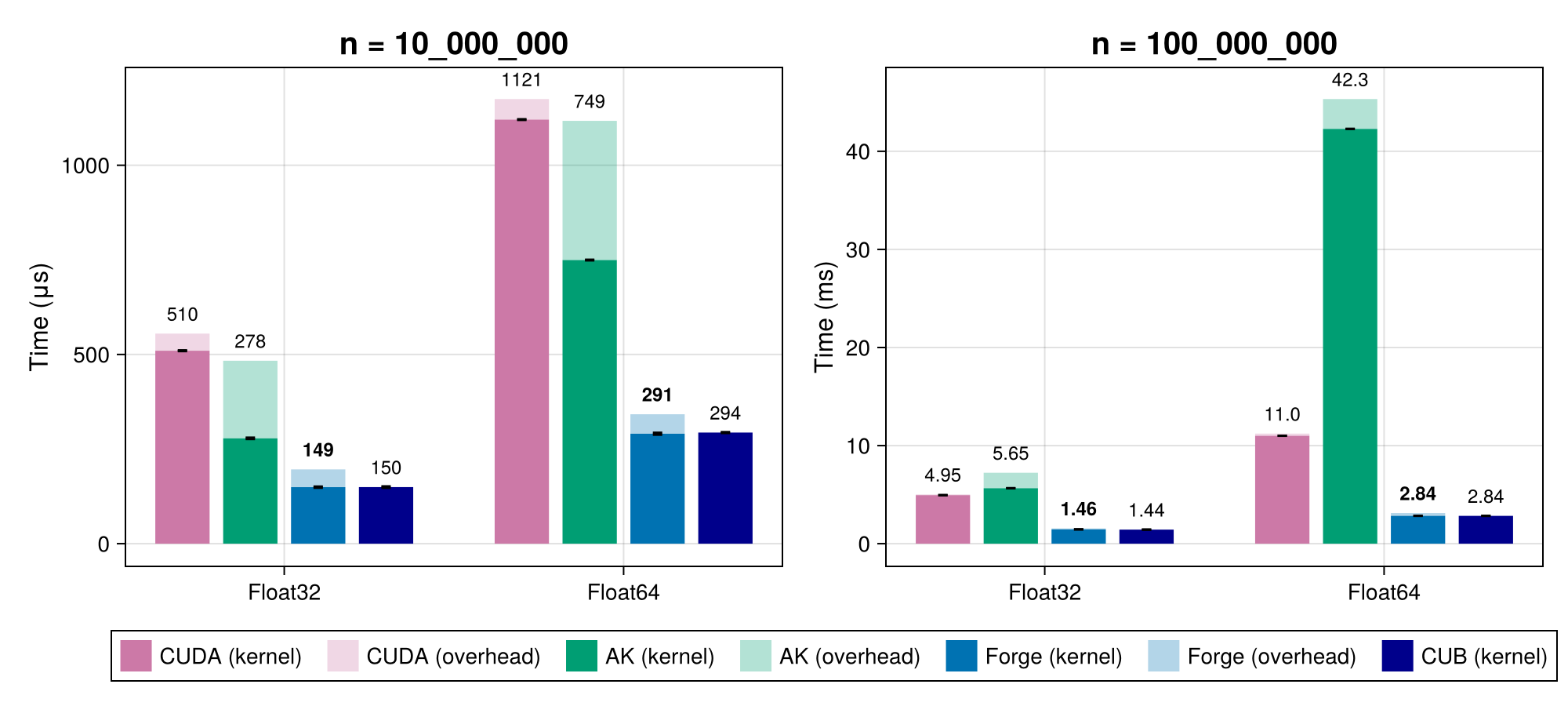}}

\caption{\textbf{Scan Benchmark Across Implementations (A40).} Same
implementations as the reduction benchmark. Each algorithm is tested with
two data types: \texttt{Float32} and \texttt{Float64}. For reference,
Kokkos achieves $0.84\,\text{ms}$, $7.4\,\text{ms}$, and
$73.4\,\text{ms}$ on \texttt{Float64} for $n = 10^7$, $10^8$, and $10^9$
respectively, representing
a $2.6\times$ overhead relative to KernelForge.jl and CUB.}
\label{fig:scan_perfs_A40}
\end{figure}

\paragraph{Scan}
Figure~\ref{fig:scan_perfs_A40} presents benchmark results for the prefix
scan primitive for input sizes $n \in \{10^7, 10^8\}$ and two data types:
\texttt{Float32} and \texttt{Float64}.

KernelForge.jl matches CUB within measurement noise across all
configurations, confirming that the decoupled lookback algorithm
(Section~\ref{subsec:scan}) achieves vendor-level throughput for both
types. CUDA.jl, which relies on a multi-launch reduction-then-scan
strategy, is $3.4\times$ slower on \texttt{Float32} and $3.9\times$ slower
on \texttt{Float64} at $n = 10^8$.

AcceleratedKernels.jl's default scan uses a sequential inter-block
accumulation pass, which becomes a bottleneck at large sizes. This is most
visible on \texttt{Float64} at $n = 10^8$, where it is $14.9\times$ slower
than KernelForge.jl. The gap widens further at $n = 10^9$, where
KernelForge.jl achieves up to $\mathbf{140\times}$ speedup over
AcceleratedKernels.jl on \texttt{Float64} (Table~\ref{tab:scan_benchmark}).
AcceleratedKernels.jl does provide a decoupled lookback variant better
suited for large inputs; we do not benchmark it here as it is non-default.

Kokkos achieves $7.4\,\text{ms}$ on \texttt{Float64} at $n = 10^8$, which is
$2.6\times$ slower relative to KernelForge.jl and CUB. Kokkos was built from
source (version 4.6.x, the latest release at time of benchmarking) with
\texttt{-DCMAKE\_BUILD\_TYPE=Release}, \texttt{-DKokkos\_ARCH\_AMPERE86=ON},
and \texttt{C++20}. The Kokkos benchmark reports zero temporary storage usage,
consistent with the fact that \texttt{Kokkos::parallel\_scan} on CUDA routes
through \texttt{Kokkos::Impl::CudaScan}---a custom implementation that does
not invoke \texttt{cub::DeviceScan} and therefore does not use the decoupled
lookback algorithm. This gap is consistent across all tested sizes ($n = 10^7$,
$10^8$, $10^9$), confirming that the performance difference reflects an
algorithmic choice rather than abstraction overhead.

\begin{figure}[htbp]
\centerline{\includegraphics[scale=0.35]{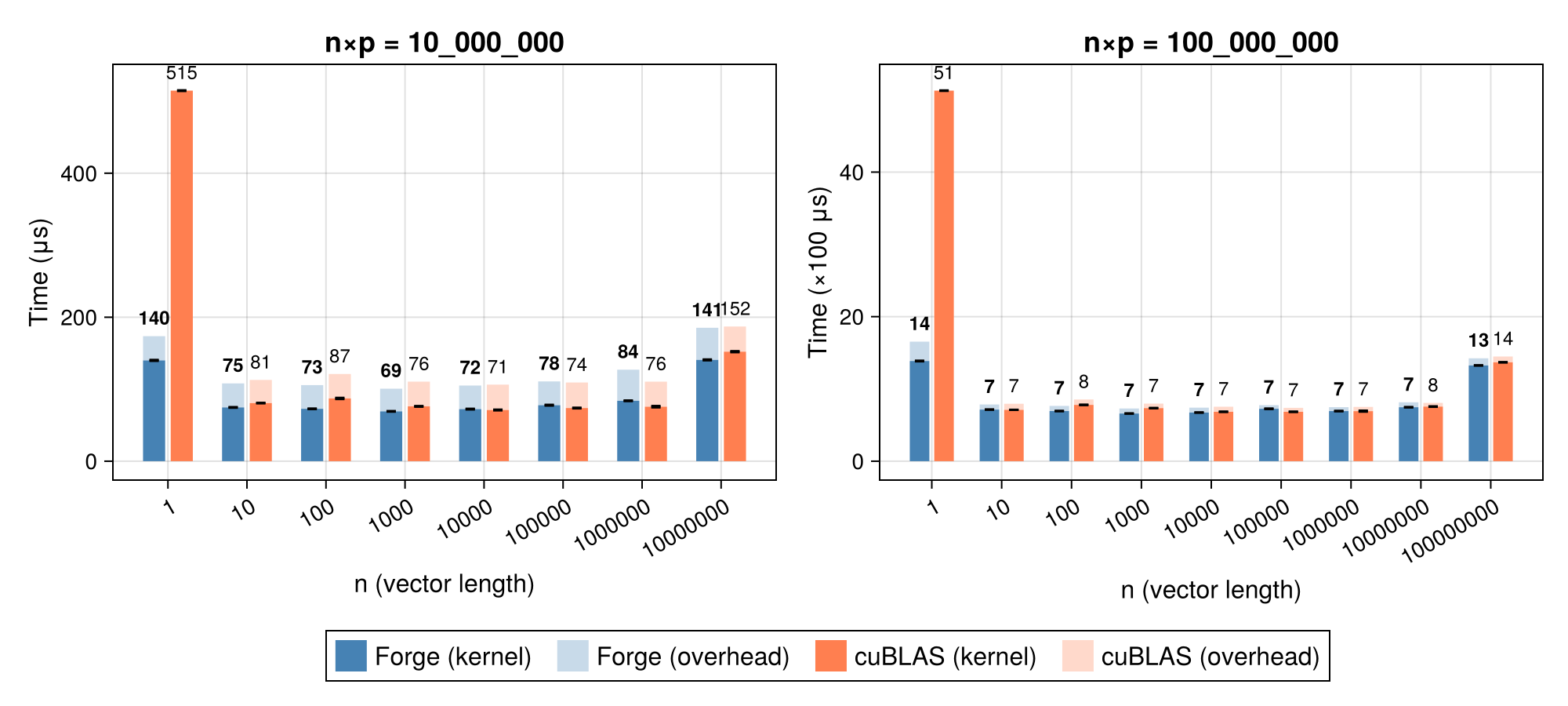}}

\caption{\textbf{Vector-Matrix Product Benchmark Across Matrix Shapes (A40).}
Throughput comparison between cuBLAS (via CUDA.jl) and KernelForge for \texttt{Float32}
vector-matrix multiplication. The total input data size $n \times p$ is fixed at $10^7$ (left)
and $10^8$ (right), with varying aspect ratios to assess performance across different
memory access patterns.}
\label{fig:vecmat_perfs_A40}
\end{figure}

\begin{figure}[htbp]
\centerline{\includegraphics[scale=0.35]{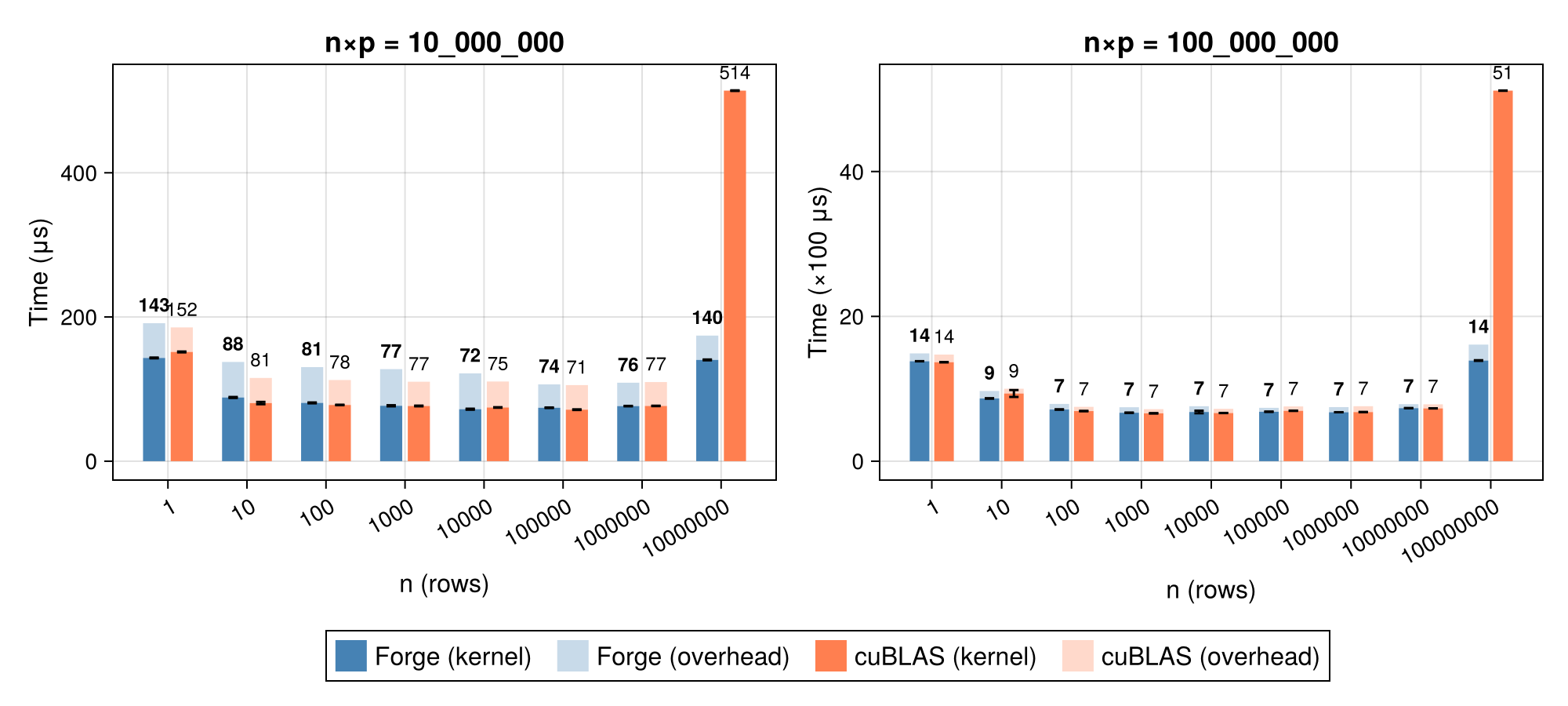}}

\caption{\textbf{Matrix-Vector Product Benchmark Across Matrix Shapes (A40).}
Throughput comparison between cuBLAS (via CUDA.jl) and KernelForge for \texttt{Float32}
matrix-vector multiplication. Same experimental setup as Figure~\ref{fig:vecmat_perfs_A40},
with total input size $n \times p$ fixed at $10^7$ (left) and $10^8$ (right).}
\label{fig:matvec_perfs_A40}
\end{figure}

\paragraph{Vector-Matrix and Matrix-Vector Operations.}
Figures~\ref{fig:vecmat_perfs_A40} and~\ref{fig:matvec_perfs_A40} compare
KernelForge.jl against cuBLAS for \texttt{Float32} vector-matrix and
matrix-vector multiplication (MatVec: row-vector $\times$ matrix; VecMat:
matrix $\times$ column-vector) with total input size $n \times p$ fixed at
$10^7$ and $10^8$, across all aspect ratios. Complete results are reported
in Tables~\ref{tab:vecmat_benchmark} and~\ref{tab:matvec_benchmark}.

For all non-degenerate aspect ratios, KernelForge.jl matches cuBLAS
throughput at $n \times p = 10^8$. At $n \times p = 10^7$, differences
within $\sim 15\%$ are observed depending on shape, with no consistent
advantage on either side. This is a direct consequence of the thread
partitioning scheme described in Section~\ref{subsec:matvec}, whose static
parameters are tuned for the A40 via the architecture dispatch system.

The degenerate cases $n = 1$ (vecmat) and $p = 1$ (matvec) reduce to a
plain memory copy, for which cuBLAS does not invoke an optimized path;
KernelForge.jl achieves up to $3.7\times$ lower kernel time on these
configurations. These cases are included for completeness only, as a
dedicated copy kernel should be preferred in practice.

At $n \times p = 10^9$, KernelForge.jl matches cuBLAS across all aspect
ratios except $(n, p) = (10^4, 10^5)$ for the matrix-vector product, where
it is $1.45\times$ slower ($9591.6\,\mu$s vs.\ $6591.4\,\mu$s). This
shape was not specifically tuned for the A40, and retuning the static
partitioning parameters may close part of this gap. However, it is also
possible that cuBLAS exploits tensor core instructions
(e.g., $4\times 1$ MMA tiles) for this shape, in which case matching its
performance without similar hardware-level access may not be achievable.

\subsection{Results on AMD MI300X}\label{sec:results_mi300x}
We evaluate KernelForge.jl on the AMD MI300X against AMDGPU.jl and
AcceleratedKernels.jl (cf.\ Tables~\ref{tab:mapreduce_benchmark_mi300x}--\ref{tab:vecmat_benchmark_mi300x}).
Unlike the NVIDIA setting, where \texttt{CUDA.@profile} isolates individual
kernel segments, ROCm timing relies on HIP events (\texttt{AMDGPU.HIP.HIPEvent}),
which measure device-side elapsed time including kernel launch overhead and any
on-device allocations. CPU-side pipeline timing is avoided, as it exhibits
substantially higher variance on ROCm---in the worst cases, the standard deviation
can exceed the mean when temporary or destination buffers are allocated within
the timed region.

To mitigate this, KernelForge.jl and AcceleratedKernels.jl are benchmarked
with pre-allocated temporary buffers---supplied via KernelForge's
\texttt{get\_allocation} interface or AcceleratedKernels.jl's \texttt{temp}
keyword argument---which both expose uniformly across all primitives.
AMDGPU.jl does not provide an equivalent facility and is therefore benchmarked
with internal allocation; its reported times should be interpreted as upper
bounds on what that library could achieve with a pre-allocation interface,
rather than as a measure of its kernel throughput alone. Consequently,
comparisons against AMDGPU.jl primarily demonstrate the practical advantage
of KernelForge's allocation-forwarding design over libraries that lack this
facility, not algorithmic superiority per se. As with CUDA.jl dispatching to
cuBLAS, AMDGPU.jl dispatches \texttt{LinearAlgebra.mul!} to \texttt{rocblas\_gemv}
internally, making it the effective rocBLAS baseline for matrix--vector and
vector--matrix primitives.

\paragraph{Mapreduce}
KernelForge.jl outperforms both baselines for \texttt{Float32} and
\texttt{UInt8} across all tested sizes. At $n = 10^9$ on \texttt{Float32},
it is $6.3\times$ faster than AMDGPU.jl and $1.4\times$ faster than
AcceleratedKernels.jl. For \texttt{UInt8}, the advantage is larger:
KernelForge.jl is $17.7\times$ faster than AMDGPU.jl and $3.6\times$
faster than AcceleratedKernels.jl at $n = 10^9$, consistent with the
vectorized memory access advantage observed on the A40. For
\texttt{UnitFloat8}, KernelForge.jl is $11.3\times$ faster than AMDGPU.jl
at $n = 10^9$, but does not consistently outperform AcceleratedKernels.jl.
On the A40, \texttt{UnitFloat8} and \texttt{UInt8} achieve comparable
throughput despite the internal promotion to \texttt{Float32}; this
parity does not hold on the MI300X, suggesting that type promotion is
handled differently by the ROCm compiler pipeline. Further tuning for
mixed-precision types on AMD hardware is left to future work.

\begin{table}[H]
\caption{Mapreduce Benchmark on AMD MI300X: Mean Kernel Time ($\mu$s) $\pm$ Std, measured via HIP events. AK: AcceleratedKernels.jl. UF8: UnitFloat8, U8: UInt8. AMDGPU times include temporary allocations; KernelForge and AK times exclude them.}
\begin{center}
\footnotesize
\resizebox{1\columnwidth}{!}{%
\begin{tabular}{|c|c|r|r|r|}
\hline
\textbf{$n$} & \textbf{Type} & \textbf{\textit{AMDGPU}} & \textbf{\textit{AK}} & \textbf{\textit{KernelForge}} \\
\hline
\multirow{3}{*}{$10^6$}
& \texttt{F32}            & $220.5 \pm 9.8$    & $52.2 \pm 5.0$    & $31.8 \pm 2.0$   \\
& \texttt{UF8$\to$F32}   & $143.9 \pm 3.8$    & $58.4 \pm 5.8$    & $72.6 \pm 12.4$  \\
& \texttt{U8}            & $110.5 \pm 3.5$    & $50.4 \pm 5.2$    & $32.7 \pm 1.8$   \\
\hline
\multirow{3}{*}{$10^7$}
& \texttt{F32}            & $125.5 \pm 3.5$    & $52.6 \pm 4.5$    & $35.7 \pm 2.0$   \\
& \texttt{UF8$\to$F32}   & $140.2 \pm 3.9$    & $92.6 \pm 4.1$    & $82.3 \pm 13.1$  \\
& \texttt{U8}            & $213.4 \pm 8.3$    & $51.2 \pm 4.1$    & $33.9 \pm 1.9$   \\
\hline
\multirow{3}{*}{$10^8$}
& \texttt{F32}            & $669.1 \pm 5.2$    & $158.4 \pm 3.7$   & $122.2 \pm 2.2$  \\
& \texttt{UF8$\to$F32}   & $646.0 \pm 4.3$    & $217.8 \pm 11.0$  & $332.8 \pm 6.4$  \\
& \texttt{U8}            & $637.2 \pm 3.1$    & $133.9 \pm 3.0$   & $54.1 \pm 1.6$   \\
\hline
\multirow{3}{*}{$10^9$}
& \texttt{F32}            & $6029 \pm 20$      & $1360 \pm 11$     & $961 \pm 5$      \\
& \texttt{UF8$\to$F32}   & $5675 \pm 16$      & $1599 \pm 16$     & $501 \pm 5$      \\
& \texttt{U8}            & $5670 \pm 10$      & $1144 \pm 3$      & $319 \pm 3$      \\
\hline
\end{tabular}}
\label{tab:mapreduce_benchmark_mi300x}
\end{center}
\end{table}

\paragraph{Scan}
KernelForge.jl matches or outperforms both baselines at $n \geq 10^7$,
with the advantage growing substantially at larger sizes. At $n = 10^8$,
KernelForge.jl is $2.7\times$ faster than AMDGPU.jl and $3.2\times$
faster than AcceleratedKernels.jl on \texttt{Float32}, and $1.9\times$
and $2.3\times$ faster respectively on \texttt{Float64}. At $n = 10^9$,
the gap widens to approximately $20\times$ over both baselines for both
types---consistent with the A40 results, where
AcceleratedKernels.jl's sequential inter-block accumulation becomes the
dominant bottleneck at large sizes. The near-identical AMDGPU.jl and
AcceleratedKernels.jl timings are expected, as AMDGPU.jl delegates its
scan implementation to AcceleratedKernels.jl internally.
At $n = 10^6$, all three implementations are within measurement noise of
each other.

\begin{table}[H]
\caption{Scan Benchmark on AMD MI300X: Mean Kernel Time ($\mu$s) $\pm$ Std, measured via HIP events. AK: AcceleratedKernels.jl. F32: Float32. F64: Float64. AMDGPU times include temporary allocations; KernelForge and AK times exclude them.}
\begin{center}
\footnotesize
\resizebox{1\columnwidth}{!}{%
\begin{tabular}{|c|c|r|r|r|}
\hline
\textbf{$n$} & \textbf{Type} & \textbf{\textit{AMDGPU}} & \textbf{\textit{AK}} & \textbf{\textit{KernelForge}} \\
\hline
\multirow{2}{*}{$10^6$}
& \texttt{F32} & $110.6 \pm 3.2$  & $41.3 \pm 2.1$  & $38.8 \pm 2.1$ \\
& \texttt{F64} & $56.0 \pm 11.8$  & $47.5 \pm 1.9$  & $52.5 \pm 9.1$ \\
\hline
\multirow{2}{*}{$10^7$}
& \texttt{F32} & $158.0 \pm 3.2$  & $141.8 \pm 2.4$  & $98.6 \pm 2.3$ \\
& \texttt{F64} & $208.5 \pm 6.7$  & $258.8 \pm 7.5$  & $146.9 \pm 13.8$ \\
\hline
\multirow{2}{*}{$10^8$}
& \texttt{F32} & $2103 \pm 10$    & $2454 \pm 26$    & $779 \pm 28$ \\
& \texttt{F64} & $2568 \pm 26$    & $3103 \pm 18$    & $1329 \pm 37$ \\
\hline
\multirow{2}{*}{$10^9$}
& \texttt{F32} & $156267 \pm 57$  & $159883 \pm 55$  & $7467 \pm 12$ \\
& \texttt{F64} & $270080 \pm 518$ & $275408 \pm 431$ & $13056 \pm 20$ \\
\hline
\end{tabular}}
\label{tab:scan_benchmark_mi300x}
\end{center}
\end{table}
\paragraph{Vector-Matrix and Matrix-Vector Operations}
KernelForge.jl achieves substantial speedups over AMDGPU.jl for
wide matrices (small $n$, large $p$). For MatVec at $n \times p = 10^8$
with $n = 10^3$, $p = 10^5$, KernelForge.jl is $3.6\times$ faster
(cf.\ Table~\ref{tab:matvec_benchmark_mi300x}); similarly large speedups
hold for VecMat at small $n$ and large $p$
(cf.\ Table~\ref{tab:vecmat_benchmark_mi300x}). However, as $n$ grows and
$p$ shrinks, AMDGPU.jl becomes competitive or faster, and for square or
tall-and-narrow shapes at $n \times p = 10^8$, KernelForge.jl offers no
consistent advantage. 

These results confirm that the algorithmic design of KernelForge.jl ports
correctly to AMD hardware and delivers competitive or superior performance
on mapreduce and scan without any AMD-specific tuning. The matrix
operation results illustrate the limits of A40-tuned parameters when
transferred to a different architecture, motivating future
per-architecture tuning work and kernel rewriting for specific aspect
ratios.

\section{Conclusion}
\label{sec:conclusion}
We presented KernelForge.jl, a Julia library demonstrating that three
properties often assumed to be mutually exclusive---performance,
flexibility, and portability---can be achieved simultaneously for
fundamental GPU parallel primitives. Through a two-layer architecture
separating portable algorithmic logic from backend-specific intrinsics,
KernelForge.jl matches vendor-optimized implementations on scan, mapreduce,
and matrix--vector operations on both NVIDIA and AMD hardware, while
supporting arbitrary element types and operators beyond what vendor libraries expose.

We regard this work as a proof of concept; several limitations remain.
Porting to additional backends (Apple Metal, Intel oneAPI) requires a
KernelIntrinsics.jl extension for each target, though the effort is
confined to the thin intrinsics layer. More critically, the static tuning
parameters were optimized for the A40 and would need retuning per
architecture; automated grid search is left to future work. Finally,
matching vendor performance on matrix--matrix products will be
significantly harder, as vendor libraries exploit tensor core instructions
that require specialization beyond warp-level primitives.

\appendix

\section{Benchmark Details on A40}\label{sec:perfs_details_A40}

Tables~\ref{tab:mapreduce_benchmark}--\ref{tab:vecmat_benchmark} report the raw kernel 
timings corresponding to the figures in Section~\ref{sec:performance}.

\begin{table}[H]
\caption{Mapreduce Kernel Benchmark on A40: Mean Kernel Time ($\mu$s) $\pm$ Std (cf.\ Figure~\ref{fig:mapreduce_perfs_A40}) for CUDA.jl, Accelerated Kernels, KernelForge and CUB.}
\begin{center}
\footnotesize
\resizebox{0.96\columnwidth}{!}{%
\begin{tabular}{|c|c|r|r|r|r|}
\hline
\textbf{$n$} & \textbf{Type} & \textbf{\textit{CUDA}} & \textbf{\textit{AK}} & \textbf{\textit{KernelForge}} & \textbf{\textit{CUB}} \\
\hline
\multirow{2}{*}{$10^6$}
& \texttt{F32}     & $10.1 \pm 0.5$ & $11.9 \pm 0.3$ & $6.1 \pm 0.2$ & $9.4 \pm 0.4$ \\
& \texttt{UF8→F32} & $7.3 \pm 0.2$  & $11.0 \pm 0.2$ & $4.9 \pm 0.2$ & $8.0 \pm 0.3$ \\
\hline
\multirow{2}{*}{$10^7$}
& \texttt{F32}     & $80.9 \pm 0.5$ & $77.0 \pm 0.4$ & $71.2 \pm 0.4$ & $75.6 \pm 0.3$ \\
& \texttt{UF8→F32} & $39.8 \pm 0.4$ & $74.4 \pm 0.4$ & $23.3 \pm 0.4$ & $25.4 \pm 3.3$ \\
\hline
\multirow{2}{*}{$10^8$}
& \texttt{F32}     & $724.9 \pm 1.2$ & $705.1 \pm 0.3$ & $679.9 \pm 1.8$ & $683.2 \pm 0.7$ \\
& \texttt{UF8→F32} & $323.7 \pm 1.0$ & $678.1 \pm 0.3$ & $178.4 \pm 0.7$ & $175.2 \pm 0.3$ \\
\hline
\multirow{2}{*}{$10^9$}
& \texttt{F32}     & $7207 \pm 9$ & $6972 \pm 2$ & $6562 \pm 2$ & $6809 \pm 2$ \\
& \texttt{UF8→F32} & $3310 \pm 5$ & $6719 \pm 1$ & $1718 \pm 3$ & $1724 \pm 3$ \\
\hline
\end{tabular}}
\label{tab:mapreduce_benchmark}
\end{center}
\end{table}

\begin{table}[H]
\caption{Scan Kernel Benchmark on A40: Mean Kernel Time ($\mu$s) $\pm$ Std (cf.\ Figure~\ref{fig:scan_perfs_A40}) for CUDA.jl, Accelerated Kernels, KernelForge and CUB.}
\begin{center}
\footnotesize
\resizebox{0.96\columnwidth}{!}{%
\begin{tabular}{|c|c|r|r|r|r|}
\hline
\textbf{$n$} & \textbf{Type} & \textbf{\textit{CUDA}} & \textbf{\textit{AK}} & \textbf{\textit{KernelForge}} & \textbf{\textit{CUB}} \\
\hline
\multirow{2}{*}{$10^6$}
& \texttt{F32} & $60.0 \pm 0.4$ & $21.2 \pm 0.2$ & $21.5 \pm 0.4$ & $20.7 \pm 1.3$ \\
& \texttt{F64} & $134.2 \pm 0.6$ & $64.7 \pm 0.5$ & $34.4 \pm 1.0$ & $38.9 \pm 0.7$ \\
\hline
\multirow{2}{*}{$10^7$}
& \texttt{F32} & $509.9 \pm 1.1$ & $278.4 \pm 1.9$ & $149.4 \pm 1.6$ & $149.5 \pm 2.0$ \\
& \texttt{F64} & $1120.6 \pm 1.0$ & $749.5 \pm 0.9$ & $290.6 \pm 2.6$ & $293.6 \pm 1.3$ \\
\hline
\multirow{2}{*}{$10^8$}
& \texttt{F32} & $4948 \pm 2.2$ & $5655 \pm 0.6$ & $1460 \pm 3.5$ & $1435 \pm 3.9$ \\
& \texttt{F64} & $11002 \pm 4.2$ & $42285 \pm 5.6$ & $2841 \pm 7.5$ & $2837 \pm 3.9$ \\
\hline
\multirow{2}{*}{$10^9$}
& \texttt{F32} & $49322 \pm 5$ & $423868 \pm 1482$ & $14553 \pm 10$ & $14287 \pm 7$ \\
& \texttt{F64} & $109724 \pm 11$ & $3944795 \pm 41$ & $28327 \pm 22$ & $28291 \pm 12$ \\
\hline
\end{tabular}}
\label{tab:scan_benchmark}
\end{center}
\end{table}

\begin{table}[H]
\caption{VecMat Kernel Benchmark on A40: Mean Kernel Time ($\mu$s) $\pm$ Std (cf.\ Figure~\ref{fig:vecmat_perfs_A40}). Rows are grouped by total input size $n \times p$.}
\label{tab:vecmat_benchmark}
\begin{center}
\resizebox{0.90\columnwidth}{!}{%
\begin{tabular}{|c|c|r|r|}
\hline
\textbf{$n$} & \textbf{$p$} & \textbf{\textit{KernelForge}} & \textbf{\textit{cuBLAS}} \\
\hline
$1$        & $10^6$  & $14.2 \pm 0.3$ & $53.5 \pm 0.2$ \\
$10$       & $10^5$  & $8.1 \pm 0.2$  & $10.1 \pm 0.3$ \\
$100$      & $10^4$  & $7.3 \pm 0.1$  & $6.6 \pm 0.1$  \\
$10^3$     & $10^3$  & $6.7 \pm 0.1$  & $11.2 \pm 0.2$ \\
$10^4$     & $100$   & $9.6 \pm 0.2$  & $12.3 \pm 0.2$ \\
$10^5$     & $10$    & $10.8 \pm 0.4$ & $7.7 \pm 0.3$  \\
$10^6$     & $1$     & $21.0 \pm 0.3$ & $18.2 \pm 0.3$ \\
\hline
$1$        & $10^7$  & $140.0 \pm 0.7$ & $514.7 \pm 0.4$ \\
$10$       & $10^6$  & $74.7 \pm 0.3$  & $80.7 \pm 0.3$  \\
$100$      & $10^5$  & $72.7 \pm 0.3$  & $87.1 \pm 0.9$  \\
$10^3$     & $10^4$  & $69.2 \pm 0.3$  & $76.1 \pm 0.4$  \\
$10^4$     & $10^3$  & $72.5 \pm 0.4$  & $71.0 \pm 0.4$  \\
$10^5$     & $100$   & $77.8 \pm 0.6$  & $73.8 \pm 0.4$  \\
$10^6$     & $10$    & $83.8 \pm 0.4$  & $75.7 \pm 0.8$  \\
$10^7$     & $1$     & $140.9 \pm 0.5$ & $152.0 \pm 0.6$ \\
\hline
$1$        & $10^8$  & $1386.5 \pm 4.3$  & $5127.4 \pm 3.0$  \\
$10$       & $10^7$  & $713.9 \pm 0.4$   & $710.5 \pm 0.3$   \\
$100$      & $10^6$  & $693.6 \pm 0.4$   & $779.0 \pm 2.0$   \\
$10^3$     & $10^5$  & $659.5 \pm 0.4$   & $734.4 \pm 1.3$   \\
$10^4$     & $10^4$  & $673.2 \pm 0.4$   & $683.4 \pm 1.0$   \\
$10^5$     & $10^3$  & $724.5 \pm 3.1$   & $683.3 \pm 0.8$   \\
$10^6$     & $100$   & $693.3 \pm 1.7$   & $692.1 \pm 6.7$   \\
$10^7$     & $10$    & $746.4 \pm 1.5$   & $755.9 \pm 3.6$   \\
$10^8$     & $1$     & $1324.8 \pm 1.7$  & $1368.1 \pm 1.6$  \\
\hline
$10$       & $10^8$  & $7104.6 \pm 1.0$    & $7083.6 \pm 2.9$    \\
$100$      & $10^7$  & $6903.7 \pm 1.3$    & $6889.4 \pm 2.2$    \\
$10^3$     & $10^6$  & $6567.2 \pm 1.8$    & $7301.5 \pm 7.4$    \\
$10^4$     & $10^5$  & $6670.5 \pm 0.9$    & $6812.7 \pm 4.4$    \\
$10^5$     & $10^4$  & $7228.4 \pm 8.4$    & $6592.3 \pm 1.6$    \\
$10^6$     & $10^3$  & $6954.2 \pm 35.5$   & $6770.5 \pm 1.9$    \\
$10^7$     & $100$   & $6927.2 \pm 47.5$   & $6830.4 \pm 4.2$    \\
$10^8$     & $10$    & $7259.4 \pm 20.2$   & $7716.6 \pm 61.8$   \\
\hline
\end{tabular}}
\end{center}
\end{table}

\begin{table}[H]
\caption{MatVec Kernel Benchmark on A40: Mean Kernel Time ($\mu$s) $\pm$ Std (cf.\ Figure~\ref{fig:matvec_perfs_A40}). Rows are grouped by total input size $n \times p$.}
\label{tab:matvec_benchmark}
\begin{center}
\resizebox{0.90\columnwidth}{!}{%
\begin{tabular}{|c|c|r|r|}
\hline
\textbf{$n$} & \textbf{$p$} & \textbf{\textit{KernelForge}} & \textbf{\textit{cuBLAS}} \\
\hline
$1$        & $10^6$  & $20.3 \pm 0.5$ & $18.5 \pm 0.4$ \\
$10$       & $10^5$  & $9.0 \pm 0.3$  & $13.0 \pm 0.2$ \\
$100$      & $10^4$  & $7.5 \pm 0.2$  & $7.2 \pm 0.2$  \\
$10^3$     & $10^3$  & $7.5 \pm 0.3$  & $10.5 \pm 0.2$ \\
$10^4$     & $100$   & $7.9 \pm 0.2$  & $11.4 \pm 0.2$ \\
$10^5$     & $10$    & $5.4 \pm 0.2$  & $4.4 \pm 0.2$  \\
$10^6$     & $1$     & $13.6 \pm 0.3$ & $53.4 \pm 0.2$ \\
\hline
$1$        & $10^7$  & $143.2 \pm 0.5$ & $151.5 \pm 0.6$ \\
$10$       & $10^6$  & $88.4 \pm 0.7$  & $80.6 \pm 1.6$  \\
$100$      & $10^5$  & $80.9 \pm 0.6$  & $78.0 \pm 0.3$  \\
$10^3$     & $10^4$  & $76.9 \pm 0.8$  & $76.6 \pm 0.5$  \\
$10^4$     & $10^3$  & $72.1 \pm 0.7$  & $74.5 \pm 0.3$  \\
$10^5$     & $100$   & $74.1 \pm 0.4$  & $71.4 \pm 0.4$  \\
$10^6$     & $10$    & $76.4 \pm 0.3$  & $76.6 \pm 0.3$  \\
$10^7$     & $1$     & $140.5 \pm 0.6$ & $514.0 \pm 0.4$ \\
\hline
$1$        & $10^8$  & $1380.4 \pm 1.1$  & $1366.8 \pm 1.7$  \\
$10$       & $10^7$  & $866.5 \pm 2.6$   & $934.8 \pm 47.6$  \\
$100$      & $10^6$  & $713.7 \pm 2.4$   & $692.1 \pm 1.1$   \\
$10^3$     & $10^5$  & $669.1 \pm 1.6$   & $661.1 \pm 1.4$   \\
$10^4$     & $10^4$  & $679.6 \pm 17.7$  & $665.2 \pm 0.9$   \\
$10^5$     & $10^3$  & $682.9 \pm 0.8$   & $696.0 \pm 1.2$   \\
$10^6$     & $100$   & $676.2 \pm 0.5$   & $678.4 \pm 0.6$   \\
$10^7$     & $10$    & $732.4 \pm 0.4$   & $729.3 \pm 0.3$   \\
$10^8$     & $1$     & $1390.0 \pm 4.2$  & $5117.6 \pm 2.1$  \\
\hline
$10$       & $10^8$  & $8785.5 \pm 12.9$   & $15063.6 \pm 186.2$ \\
$100$      & $10^7$  & $7061.4 \pm 10.5$   & $6814.6 \pm 4.5$    \\
$10^3$     & $10^6$  & $6584.1 \pm 8.5$    & $6686.6 \pm 10.1$   \\
$10^4$     & $10^5$  & $9591.6 \pm 20.2$   & $6591.4 \pm 24.0$   \\
$10^5$     & $10^4$  & $6902.6 \pm 4.5$    & $6794.1 \pm 11.6$   \\
$10^6$     & $10^3$  & $6620.4 \pm 1.9$    & $6843.1 \pm 2.9$    \\
$10^7$     & $100$   & $6707.5 \pm 3.1$    & $6735.0 \pm 4.1$    \\
$10^8$     & $10$    & $7278.7 \pm 0.9$    & $7259.3 \pm 0.7$    \\
\hline
\end{tabular}}
\end{center}
\end{table}

\newpage

\newpage 

\section{Benchmark of whole pipeline on AMD MI300X}\label{sec:perf_amd}

\begin{table}[H]
\caption{MatVec Benchmark on AMD MI300X: Mean Kernel Time ($\mu$s) $\pm$ Std, measured via HIP events. Rows grouped by total input size $n \times p$. AMDGPU times include temporary allocations; KernelForge times exclude them.}
\label{tab:matvec_benchmark_mi300x}
\begin{center}
\resizebox{1\columnwidth}{!}{%
\begin{tabular}{|c|c|r|r|}
\hline
\textbf{$n$} & \textbf{$p$} & \textbf{\textit{KernelForge}} & \textbf{\textit{AMDGPU}} \\
\hline
$1$        & $10^7$  & $263.7 \pm 2.9$      & $53558 \pm 430$       \\
$10$       & $10^6$  & $388.6 \pm 3.1$      & $5681.5 \pm 39.6$    \\
$100$      & $10^5$  & $262.1 \pm 2.9$      & $835.6 \pm 7.5$      \\
$10^3$     & $10^4$  & $133.8 \pm 3.9$      & $148.0 \pm 1.9$      \\
$10^4$     & $10^3$  & $189.3 \pm 3.4$      & $174.5 \pm 116.8$    \\
$10^5$     & $100$   & $30.9 \pm 1.2$       & $115.9 \pm 8.7$      \\
$10^6$     & $10$    & $33.7 \pm 1.4$       & $29.3 \pm 14.5$      \\
$10^7$     & $1$     & $43.7 \pm 1.9$       & $1021.3 \pm 6283.3$  \\
\hline
$1$        & $10^8$  & $2797.6 \pm 10.3$    & $737748 \pm 4006$     \\
$10$       & $10^7$  & $4440.5 \pm 9.8$     & $75250 \pm 386$       \\
$100$      & $10^6$  & $1961.0 \pm 4.8$     & $10572 \pm 44$        \\
$10^3$     & $10^5$  & $307.3 \pm 43.2$     & $1096.5 \pm 3.5$     \\
$10^4$     & $10^4$  & $251.6 \pm 7.1$      & $141.1 \pm 2.1$      \\
$10^5$     & $10^3$  & $155.1 \pm 2.0$      & $163.0 \pm 8.6$      \\
$10^6$     & $100$   & $142.3 \pm 2.4$      & $170.4 \pm 17.0$     \\
$10^7$     & $10$    & $201.6 \pm 3.7$      & $159.8 \pm 36.7$     \\
$10^8$     & $1$     & $285.7 \pm 4.4$      & $13218 \pm 23618$    \\
\hline
$10$       & $10^8$  & $43906 \pm 171$      & $752899 \pm 4023$    \\
$100$      & $10^7$  & $19112 \pm 39$       & $105736 \pm 403$     \\
$10^3$     & $10^6$  & $2140.2 \pm 4.9$     & $10869 \pm 23$       \\
$10^4$     & $10^5$  & $1410.7 \pm 21.4$    & $1294.3 \pm 8.6$     \\
$10^5$     & $10^4$  & $1412.5 \pm 17.1$    & $1086.2 \pm 30.9$    \\
$10^6$     & $10^3$  & $1219.1 \pm 13.7$    & $1011.9 \pm 42.7$    \\
$10^7$     & $100$   & $1299.1 \pm 7.1$     & $1162.9 \pm 17.0$    \\
$10^8$     & $10$    & $1872.9 \pm 22.0$    & $27329 \pm 233531$   \\
\hline
\end{tabular}}
\end{center}
\end{table}

\begin{table}[H]
\caption{VecMat Benchmark on AMD MI300X: Mean Kernel Time ($\mu$s) $\pm$ Std, measured via HIP events. Rows grouped by total input size $n \times p$. AMDGPU times include temporary allocations; KernelForge times exclude them.}
\label{tab:vecmat_benchmark_mi300x}
\begin{center}
\resizebox{1\columnwidth}{!}{%
\begin{tabular}{|c|c|r|r|}
\hline
\textbf{$n$} & \textbf{$p$} & \textbf{\textit{KernelForge}} & \textbf{\textit{AMDGPU}} \\
\hline
$1$        & $10^7$  & $36.9 \pm 1.1$      & $448.4 \pm 17.3$    \\
$10$       & $10^6$  & $29.8 \pm 1.3$      & $133.2 \pm 16.8$    \\
$100$      & $10^5$  & $27.8 \pm 1.2$      & $108.9 \pm 8.3$     \\
$10^3$     & $10^4$  & $30.5 \pm 5.0$      & $82.8 \pm 2.3$      \\
$10^4$     & $10^3$  & $28.5 \pm 1.2$      & $103.0 \pm 3.1$     \\
$10^5$     & $100$   & $48.9 \pm 2.0$      & $137.4 \pm 1.6$     \\
$10^6$     & $10$    & $52.6 \pm 8.9$      & $98.8 \pm 3.9$      \\
$10^7$     & $1$     & $56.9 \pm 10.2$     & $104.8 \pm 2.3$     \\
\hline
$1$        & $10^8$  & $283.6 \pm 3.5$     & $4740.0 \pm 156.5$  \\
$10$       & $10^7$  & $145.6 \pm 2.7$     & $686.1 \pm 94.7$    \\
$100$      & $10^6$  & $162.5 \pm 8.0$     & $169.4 \pm 13.7$    \\
$10^3$     & $10^5$  & $137.2 \pm 12.5$    & $175.0 \pm 6.8$     \\
$10^4$     & $10^4$  & $131.9 \pm 12.3$    & $164.4 \pm 12.6$    \\
$10^5$     & $10^3$  & $130.0 \pm 1.9$     & $185.8 \pm 1.7$     \\
$10^6$     & $100$   & $187.3 \pm 3.2$     & $148.2 \pm 3.4$     \\
$10^7$     & $10$    & $156.0 \pm 2.7$     & $158.6 \pm 2.5$     \\
$10^8$     & $1$     & $212.4 \pm 2.8$     & $232.1 \pm 3.4$     \\
\hline
$10$       & $10^8$  & $1410.6 \pm 12.8$   & $6877.0 \pm 82.9$   \\
$100$      & $10^7$  & $1452.2 \pm 29.6$   & $1247.6 \pm 30.2$   \\
$10^3$     & $10^6$  & $1180.1 \pm 28.3$   & $1495.2 \pm 14.5$   \\
$10^4$     & $10^5$  & $1179.4 \pm 42.0$   & $1079.9 \pm 8.5$    \\
$10^5$     & $10^4$  & $1040.6 \pm 44.0$   & $1112.3 \pm 4.5$    \\
$10^6$     & $10^3$  & $1176.1 \pm 7.9$    & $1640.3 \pm 4.3$    \\
$10^7$     & $100$   & $1212.6 \pm 29.1$   & $1159.2 \pm 53.8$   \\
$10^8$     & $10$    & $1411.9 \pm 13.1$   & $1242.1 \pm 66.4$   \\
\hline
\end{tabular}}
\end{center}
\end{table}

\bibliographystyle{IEEEtran}
\bibliography{biblio}

\end{document}